\documentclass[twocolumn,showpacs,preprintnumbers,amsmath,amssymb,prl,superscriptaddress,floatfix]{revtex4}
\usepackage{color}
\usepackage{graphicx}% Include figure files
\usepackage{dcolumn}% Align table columns on decimal point
\usepackage{bm}% bold math
\usepackage[caption=false]{subfig}
\usepackage{placeins}
\usepackage{multirow}
\usepackage{amsmath}
\usepackage{multirow}

\begin{document}
%\preprint{APS/123-QED}
%: local damage can lead to total system collapse
\title{Spatially localized attacks on interdependent networks: the existence of a finite critical attack size}
%Spatially localized attacks with critical damage size on interdependent networks: critical attack size}
%Spatially localized attacks on interdependent networks: global collapse can be triggered by minute local failures}
%Spatially localized attacks with critical damage size on interdependent networks
%{Minute local attacks on spatially embedded interdependent networks may lead to system collapse}% Force line breaks with \\
% \email{bereziny@google.com}

\author{Yehiel Berezin}
\affiliation{Department of Physics, Bar Ilan University, Ramat Gan 52900, Israel}

\author{Amir  Bashan}%
\affiliation{Department of Physics, Bar Ilan University, Ramat Gan 52900, Israel}

\author{Michael M. Danziger}%
\affiliation{Department of Physics, Bar Ilan University, Ramat Gan 52900, Israel}

\author{Daqing Li}
\affiliation{School of Reliability and Systems Engineering, Beihang University, Beijing 100191, China}

\author{Shlomo Havlin}%
\affiliation{Department of Physics, Bar Ilan University, Ramat Gan 52900, Israel}

%Use showkeys class option ifkeyword

%\address{

\date{\today}% It is always \today, today,

\begin{abstract}
Many real world complex systems such as infrastructure, communication and transportation networks are embedded in space, where entities of one system may depend on entities of other systems.
These systems are subject to geographically localized failures due to malicious attacks or natural disasters.
Here we study the resilience of a system composed of two interdependent spatially embedded networks to localized geographical attacks.
We find that if an attack is larger than a finite (zero fraction of the system) critical size, it will spread through the entire system and lead to its complete collapse. If the attack is below the critical size, it will remain localized.
In contrast, under random attack a finite fraction of the system needs to be removed to initiate system collapse.
We present both numerical simulations and a theoretical approach to analyze and predict the effect of local attacks and the critical attack size.
Our results demonstrate the high risk of local attacks on interdependent spatially embedded infrastructures and can be useful for designing more resilient systems.
%We find that initial local attacks above a critical size, which does not increase with system size, spread through the entire system and lead to a total collapse of the system, while below the critical size the damage does not spread.
%The critical size is a zero fraction of the system size, contrary to the case of a random attack where a finite fraction is needed to initiate system collapse.
%The critical zero size is depend on the system dependency length $r$.
%(modeled as square lattices)
%The critical zero size is depend on the system dependency length $r$.
\end{abstract}

%\pacs{89.75.-k}{Complex systems}
%\pacs{05.40.-a}{Fluctuation phenomena, random processes, noise, and Brownian
%motion}
%\pacs{89.60.Gg}{Environmental studies}

%% Classification Scheme.
%\keywords{Suggested keywords} Use showkeys class option if keyword
% display desired
\maketitle
%\section{\label{sec:level1}Introduction}
Modern critical infrastructures are embedded in space and have extensive interdependencies.
Entities in one network (e.g., power generation/distribution, communications, transportation etc.) are dependent upon entities in another and failures in one network can trigger failures in another.
It has been shown that these dependencies lead to substantially decreased robustness and even abrupt
first order transitions which are absent in isolated
networks~\cite{rinaldi-ieee2001,peerenboom-proceedings2001,rosato-criticalinf2008,buldyrev-nature2010,leichtdsouza2009,parshani-prl2010,bashan-naturephysics2013,vespignani-nature2010,gao-prl2011,hu-pre2011,brummitt-pnas2012,gao-naturephysics2012,son-epl2012,zhao-jstatmech2013,baxter-prl2012,cellai-arxiv2013}.
For spatially embedded interdependent networks under {\it random} attack, it was shown that if the maximal dependency link length is above a critical value a new kind of abrupt collapse occurs,
characterized by a uniform spreading process~\cite{wei-prl2012,bashan-naturephysics2013}. However, a purely {\it random} failure of a finite fraction of nodes in a very large network can be unrealistic.
A more realistic scenario is a failure of a group of neighboring nodes due to a natural disaster like the 2011 T\={o}hoku earthquake and tsunami or due to a malicious attack affecting all networks in a given region (e.g., a nuclear strike) or only certain infrastructures (e.g., an electromagnetic pulse
or chemical/biological attack). The resilience of a system of interdependent networks to an attack of this sort, which we call ``localized attack,'' has not been addressed before.

\begin{figure}[ht*]
  \begin{center}
  \includegraphics[width=\linewidth,trim=13mm 13mm 15mm 8mm,clip]{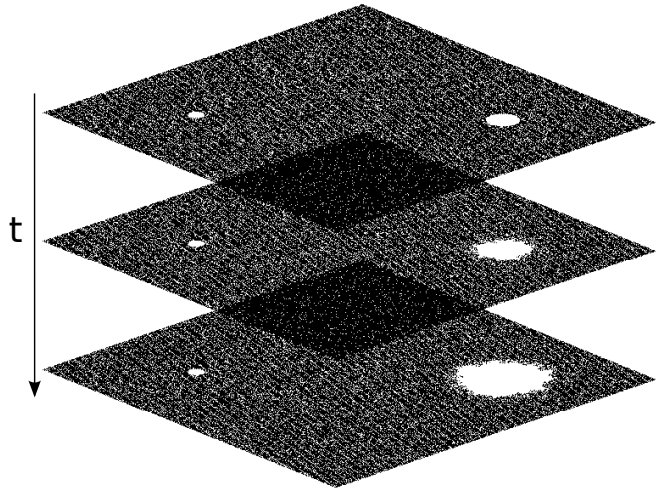}
  \end{center}
  %\includegraphics[]{}\\
  \caption{Propagation of local damage throughout the network. Under the same system conditions, the hole on the right is above the critical size $r_h^c$ and spreads while the hole on the left is below $r_h^c$ and remains almost the same size.}
  \label{fig:prop-diagram}
\end{figure}

We show here that there exists a critical damage size with radius $r_h^c$, above which localized  geographical damage will spread
and destroy the whole system and below which it will remain localized (see Fig. \ref{fig:prop-diagram}).
This critical size is determined
solely by intensive system quantities and thus, in contrast to random failures, constitutes a zero-fraction of the system in the large system limit, $N\rightarrow\infty$.
%We find that there are distinct stable, unstable and metastable phases depending on the average degree $\langle k \rangle$ and the maximal length of dependency links $r$.  In the metastable phase, there exists a critical damage size, with radius $r_h^c$,  above which a localized geographical damage will spread and destroy the whole sy0stem, while below $r_h^c$ it remains localized. This critical size is determined solely by intensive system quantities and thus constitutes a zero-fraction of the system in large systems limit, $N\rightarrow\infty$.

The resilience of single complex networks to random attacks~\cite{albert-nature1999,cohen-prl2000,callaway-prl2000} and malicious attacks targeting nodes with special topological properties~\cite{callaway-pre2001,gallos-compscience2006} has been studied.
%The resilience of a complex network to random attack or to malicious attacks based on targeting special nodes (by degree, betweenness etc.) has been studied extensively in recent years.\cite{albert-nature1999,cohen-prl2000,callaway-pre2001,gallos-compscience2006}
%It has been shown that the robustness of the network under such an attack is highly dependent on it topology.
Resilience to geographic localized attacks has been studied on specific single networks~\cite{neumayer-milcom2008,agarwal-milcom2010,agarwal-infocom2011,neumayer-ieee2011},
but a general theoretical approach of such attacks is currently missing.
In particular,  the effects of cascading failures due to interactions between networks has not been evaluated with respect to localized attacks even though the positive feedback caused by interdependencies has been shown to have catastrophic consequences such as the 2003 Italian blackout which resulted from a localized failure in a system of interdependent networks~\cite{rosato-criticalinf2008}.

The introduction of a percolation framework for random coupled networks~\cite{buldyrev-nature2010,leichtdsouza2009,parshani-prl2010} brought attention to many other properties of coupled networks.
Examples include the study of transport~\cite{morris-prl2012}, epidemic spreading~\cite{serrano-pre2012}, diffusion~\cite{aguirre-naturephysics2013}, suppressing cascading loads~\cite{brummitt-pnas2012}, designing robust coupled networks~\cite{schneider-scireports2013} and dynamical transitions in coupled networks~\cite{radicchi-preprint2013}.

%{\color{blue}
%Coupled transport \cite{morris-prl2012} percolation based on epidemic spreading \cite{son-epl2012} Epidemic spreading on interconnected networks \cite{serrano-pre2012} diffusion dynamics on multiplex %networks \cite{gomez-prl2013} Successful strategies for competing networks \cite{aguirre-naturephysics2013} Suppressing cascades of load in interdependent networks \cite{brummitt-pnas2012} Towards %designing robust coupled networks \cite{schneider-scireports2013} Abrupt transition in the structural formation of interconnected networks \cite{radicchi-preprint2013}
%}

%Previous research on random networks has shown that the introduction of dependencies between networks can lead to severely decreased robustness and even abrupt first order transitions \cite{leichtdsouza2009,parshani-prl2010,buldyrev-nature2010,vespignani-nature2010,gao-prl2011,hu-pre2011,gao-naturephysics2012,brummitt-pnas2012,zhao-jstatmech2013}. For spatially embedded interdependent networks under random attack, it was shown that if the maximal dependency link length is above a critical value a new kind of first-order transition, characterized by a uniform spreading process, occurs~\cite{wei-prl2012,bashan-naturephysics2013}. However, no study on interdependent networks has focused on percolation due to spatially localized attacks.

%In this transition, a random attack on the network creates a small hole which spreads from iteration to iteration and destroys the entire network.

In this Letter, we study the new phenomenon of localized attacks on interdependent spatially embedded networks.
%For the case of a single spatially embedded network, local damage will only affect its immediate neighborhood.
We find that even though the damage, connectivity and dependency links are all highly localized, a small local attack (independent of system size) can spread and destroy the entire network.
%We also find that even in a network that is not diluted ($k=4$), a small hole can initiate a cascade which can destroy the entire network.
We show that the system will fail if a geographically local attack is greater than a critical size which is a zero-fraction of the system size.
These results have profound implications for the role of network topology in the design of resilient infrastructures.

\begin{figure}[ht*]
\centering
 \subfloat{\includegraphics[width=0.5\linewidth,height=0.5\linewidth]{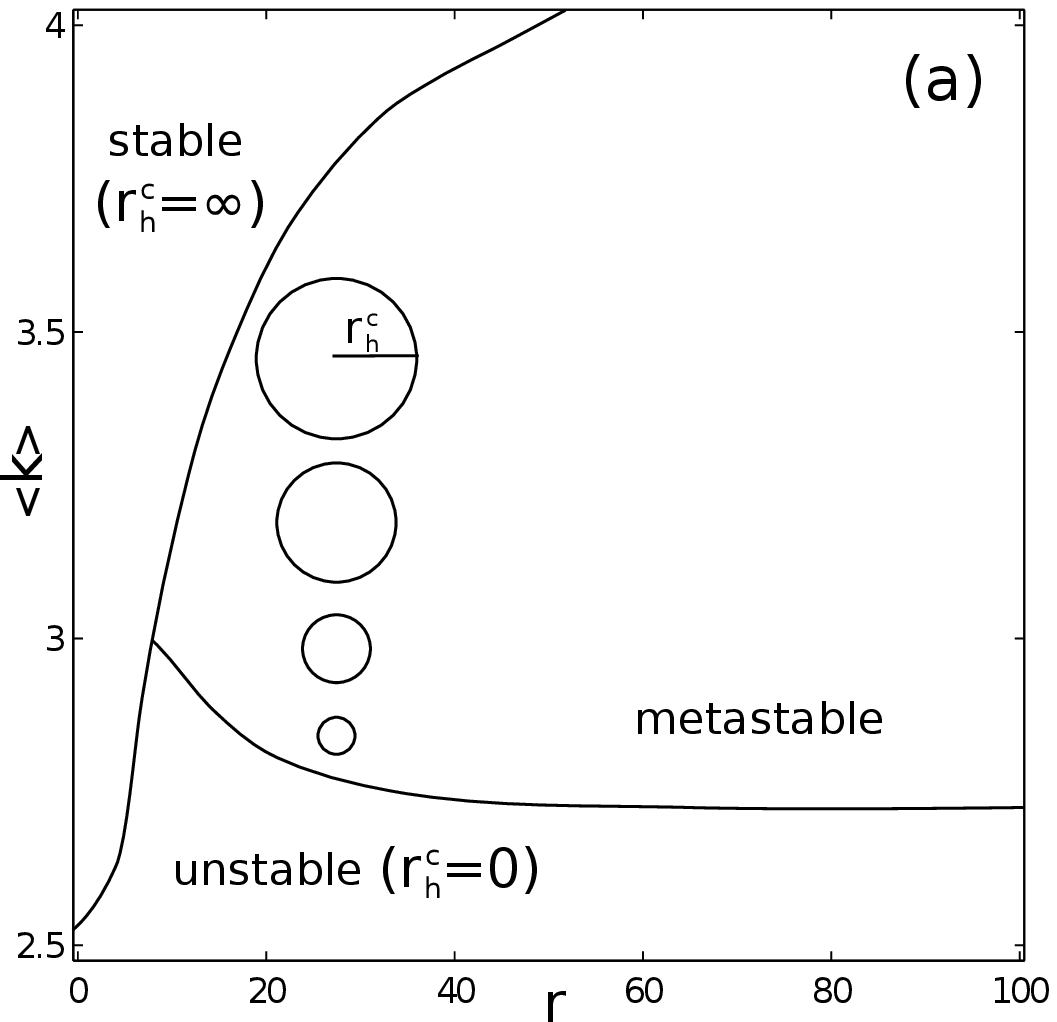}\label{fig:phase-diagram}}
 \subfloat{\includegraphics[trim=0 -8mm 0 0 ,width=0.48\linewidth,height=0.5\linewidth]{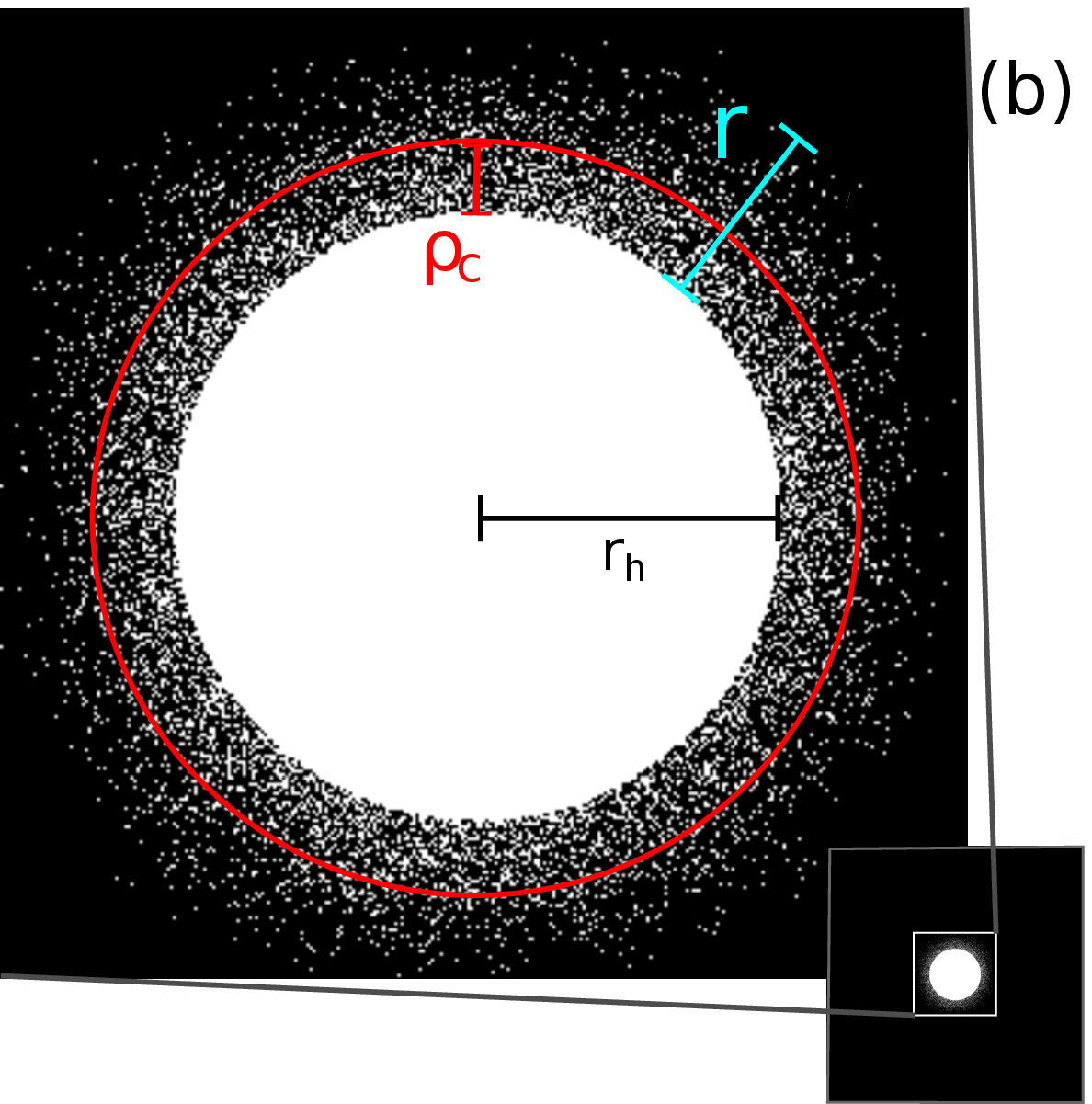}\label{fig:gradient}}%{strip.eps}
   \caption{Phase diagram of the problem. (a) Depending on $\langle k \rangle$ and $r$, the system is either stable, unstable or metastable.
   The circles shown in the metastable region illustrate the sensitivity of the critical attack size that leads to system collapse in the metastable region. For precise critical sizes for the whole metastable region see Fig. \ref{simANDmodel} below.
  (b) Demonstration of the theoretical considerations. Near the edge of a hole, the survival probability of a node increases with the distance from the edge.  The parameter $\rho_c$ denotes the distance from the edge of the hole at which the occupation probability is equal to $p_c\approx 0.5927$.
  %Even when $\rho<\rho_c$ and the concentration is below $p_c$, the functional nodes form clusters of size $\xi_{<}$ which need to be smaller than $\rho_c$ in order to fully separate.
  In the case illustrated here, the clusters have ample room to fall off and the damage will propagate and destroy the whole system, even though the relative size of the hole is small. }
\end{figure}

We model spatially embedded networks by assuming two square lattices $A$ and $B$ with periodic boundary conditions and overlaying them both on the same Cartesian plane.
Each node in network $A$ is dependent upon a node in network $B$ (and vice versa) which is chosen at random from all of the nodes within a radius $r$.
If a node in $A$ is dependent on a node in $B$, the failure of the node in $B$ will cause the node in $A$ to fail immediately and vice versa.
These dependency relationships are taken to be mutual to prevent a single failure from propagating through the entire system~\cite{gao-naturephysics2012}.

The interdependent networks are then diluted from degree $k=4$ to a lower average degree.
This is accomplished by removing a random fraction $1-p$ of the nodes from the system, along with the links that are attached to them.
This removal triggers a cascade which leaves the average degree $\langle k \rangle$ lower than its value after site dilution of $1-p$ on a single lattice. %This removal can also be interpreted as a random uniform probability of failure.
Our motivation in reducing the degree from $4$ is based on empirical studies of the power grid which have shown a characteristic degree of $ \langle k \rangle \approx 3$~\cite{hines-proceedings2010}.

We examine the effects of localized geographical damage of characteristic size $r_h$ for systems with different values of $r$ and $\langle k \rangle$.  We model this damage by removing a hole of radius $r_h$ from a random location in network $A$.
This triggers a cascade in which the nodes in $B$ which depend on the removed nodes fail, triggering further losses as more nodes in $B$ get cut off from the largest connected component.
The percolative damage in $B$ triggers further damage in $A$ due to the dependencies between the networks.
This process is continued iteratively until no more nodes fail. At the end of this cascade, the system is categorized as functional or non-functional depending on whether a finite-fraction largest connected component remains or not.

For every system with a given $r$ (maximal dependency link length) and $\langle k \rangle$ we find that there is a critical damage size $r_h^c$ below which the system remains intact and above which the damage propagates throughout the system and destroys it.
%We show numerically and theoretically that the value of $r_h^c$ is not dependent on system size.
Furthermore, we discovered three distinct phases in this system according to which the $k\text{--}r$ plane can be divided into three distinct regions as shown in Fig.~\ref{fig:phase-diagram}.
% \begin{figure}
% %,trim=5mm -5mm 8mm 5mm,clip=true
%  \subfloat[$r_h^c(\langle k \rangle)$]{\includegraphics[width=0.5\linewidth]{critical_hole_vs_k_discrete_r.eps}\label{fig:rhck}}
%    \subfloat[$r_h^c(r)$]{\includegraphics[width = 0.5\linewidth]{critical_hole_vs_system_r_discrete_k.eps}\label{rhcr}} \\
% \caption{The critical attack size as function of average degree ($\langle k \rangle$) as well as the system dependency length ($r$).}
% \label{fig:rhc-plots}
% \end{figure}

\begin{figure}[ht]
 \includegraphics[width=0.93\linewidth]{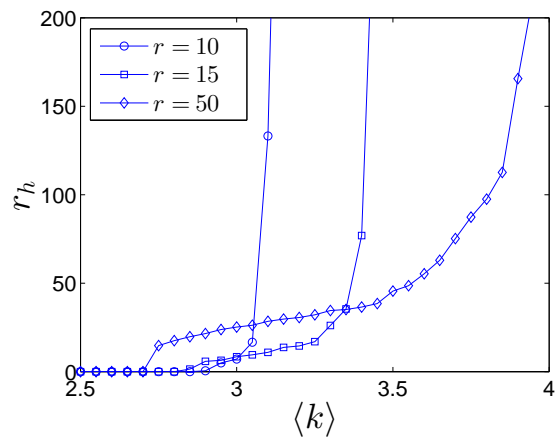}
 \caption{The critical attack size as function of average degree $\langle k \rangle$ for several $r$ values.
 The curves represent moving along vertical lines from bottom to top in Fig.~\ref{simANDmodel} (the shown circles in Fig.~\ref{fig:phase-diagram}) and show how $r_h^c$ varies with $\langle k \rangle$.
 For small $r$ the metastable state disappears and $r_h^c$ jumps from zero to infinity while for larger $r$ there is an intermediate regime where $r_h^c$ increases gradually.}
 %As $r$ decreases, the effect of $r_h$ is less pronounced and the system nears its behavior at $r<r_c$ in which there is no metastable region, see Fig. \ref{simANDmodel} and Ref \cite{wei-prl2012}.}
 \label{fig:rhck}
\end{figure}

\begin{figure}[ht]
 \includegraphics[width=0.95\linewidth]{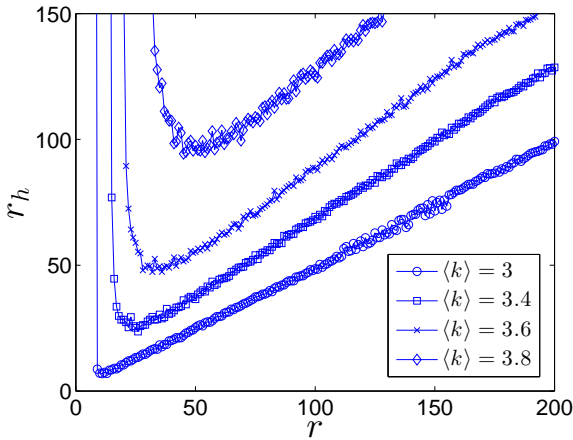}
 \caption{The critical attack size as function of system dependency length $r$ for several $\langle k \rangle$ values.
 The curves represent the value of $r_h^c$ when moving along horizontal lines in  Fig.~\ref{simANDmodel} from right to left.
 The critical damage size $r_h^c$ takes a minimal value and the system is most susceptible to small local attacks when $r$ is near the stable phase.  }
 \label{fig:rhcr}
\end{figure}
%This seemingly unintuitive result is explained below.
In the stable region, no matter how large $r_h$ is (as long as it is finite) the damage will remain localized and the system will stay intact.
In the unstable region, the system spontaneously collapses even with $r_h = 0$ (no localized attack).  This is because in this region large holes develop spontaneously due to percolation and overwhelm the system~\cite{wei-prl2012}.
The intermediate region is metastable.  Without the removal of a hole or the removal of a hole smaller than $r_h^c$, the system remains intact.
However, if a hole of size $\geq r_h^c$ is removed, it will begin a cascade that will eventually destroy the entire system.
This metastability is analogous to the well known supercooling property of water in which water can be cooled well below its freezing point and remain in the liquid phase until a disturbance of some sort triggers crystallization and it turns to ice~\cite{debenedetti-physicstoday2003}.

In Figs.~\ref{fig:rhck} and~\ref{fig:rhcr} we show, based on numerical simulations, how the critical damage size $r_h^c$ changes with respect to $r$ and $\langle k \rangle$.
In Fig.~\ref{fig:rhck}, we see that for low $\langle k \rangle$, $r_h^c$ is very small.  For larger $\langle k \rangle$, we see that $r_h^c$ increases dramatically at a certain $\langle k \rangle$ value.
The jump occurs at larger $\langle k \rangle$ values for larger $r$ values.
In Fig.~\ref{fig:rhcr}, we find that the minimum of $r_h^c$ is found near the lowest $r$ of the metastable region, making it most susceptible to local damage.

Since the metastable region spreads over a wide range of realistic values of $r$ and $\langle k \rangle$, it is of great interest to understand how this transition takes place and to predict the value of $r_h^c(r,k)$. To present a theoretical understanding of this phenomenon, we first consider in detail the chain of events triggered by the localized geographical damage.
When a hole of $r_h$ is removed from $A$, it can directly disable nodes in $B$ up to a distance $r$ from its edge.
The probability that a given node in $B$ was dependent on one of the removed nodes is highest at the edge of the hole and decreases linearly until it equals zero at distance $r$.
This creates a lattice concentration gradient in the form of an annulus of width $r$ surrounding the removed hole, see Fig.~\ref{fig:gradient}.  %By calculating the value of that gradient, we can predict whether a hole with radius $r_h$ will propagate or not.
Taking $\rho$ as the distance from the edge of the hole, the gradient of occupation probability following an attack can be evaluated as
\begin{equation}\label{eq:p}
 p(\rho , r , r_h , \langle k \rangle) = p_s(\langle k \rangle)\frac{I(r_h,r,\rho)}{\pi r^2}
\end{equation}
where $p_s(\langle k \rangle)$ is the system-wide occupation concentration and $I(r_h,r,\rho)$ is the standard formula for the area of intersection of two circles of radius $r$ and $r_h$ with centers located a distance $\rho + r_h$ from each other.
For a given set of system parameters ($r$, $r_h$, $\langle k \rangle$) we can set $p = p_c$ in Eq.~(\ref{eq:p}) and solve for $\rho$.
If a solution in the region of interest ($0<\rho<r$) exists, it corresponds to a distance $\rho_c$ at which the lattice concentration will be equal to its critical value.
The existence of such a point is a necessary but not sufficient condition for the hole to propagate.
Below $p_c$, the lattice does not spontaneously disintegrate but rather forms clusters of characteristic size $\xi_{<}(p)$, which diverges at $p_c$~\cite{bunde1991fractals,staufferaharony}.

% \begin{figure}
% %  \includegraphics[width=\linewidth]{grad_illustration_r_50_h_100.eps}%{strip.eps}
% \caption{}\label{fig:strip}
% \end{figure}

\begin{figure}[ht]
  \centering
  \begin{picture}(100,100)
  \put(-70,-10){\includegraphics[width=\linewidth,trim=23mm 33mm 25mm 38mm,clip]{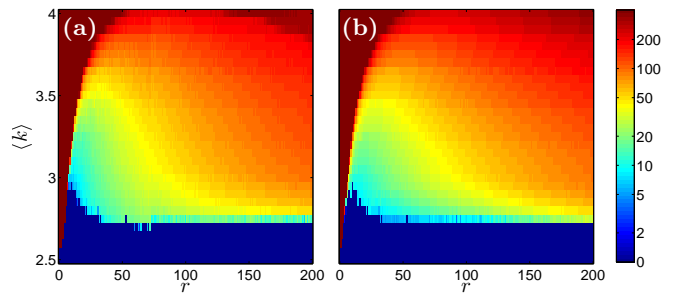}}
  \put(-51,95){\color{white}\textbf{(a)}}
  \put(51,95){\color{white}\textbf{ (b)}}
\end{picture}

  %{critical_hole_vs_system_r_vs_keff_sim_and_anal_rmax_200_iso_xi_c_0.08_big.eps}

  %\includegraphics[]{}\\
  \caption{The critical attack size, $r_h^c$ (represented by the color), as a function of $\langle k \rangle$ and the system dependency length $r$.
  (a) Simulation. We use a binary search algorithm to find the critical radius size where the local attack spreads through the entire system.
   %The black lines shows the ``isolevels''\textemdash points at which the critical hole is of the same size for different ($k,r$) values.
  (b) Model. The critical size is calculated as the smallest value of $r_h$ for which Eq.~(\ref{eq:selfcons}) has a self-consistent solution.}
  \label{simANDmodel}
\end{figure}
%
% \begin{figure}[ht]
%   \begin{center}
%   %\includegraphics[width=\textwidth,height = 0.4\textwidth]{critical_hole_vs_system_r_vs_p_sim_and_anal_rmax_200_iso_xi_c_0.15_lowRes.eps}
%   \includegraphics[width=\linewidth]{critical_hole_vs_system_r_vs_keff_sim_and_anal_rmax_200_iso_xi_c_0.8_p_c_0.85_big_v1.eps}
% %{critical_hole_vs_system_r_vs_keff_sim_and_anal_rmax_200_iso_xi_c_0.08_big.eps}
%
%   \end{center}
%   % Requires \usepackage{graphicx}
%   %\includegraphics[]{}\\
%   \caption{The critical attack size, $r_h^c$ (shown in the colored bar), as a function of $\langle k \rangle$ and the system dependency length $r$.
%   (a) Simulation. We use a binary search algorithm to find the critical radius size where the local attack spreads along the entire system.
%    %The black lines shows the ``isolevels''\textemdash points at which the critical hole is of the same size for different ($k,r$) values.
%   (b) Model. The critical size is calculated as the smallest value of $r_h$ for which Eq. (\ref{eq:selfcons}) has a self-consistent solution.}
%   \label{simANDmodel}
% \end{figure}

Hence the critical region $0<\rho<\rho_c$ needs to be wide enough for clusters of size $\xi_{<}(p)$ to form and break away.% as is illustrated in Fig. \ref{fig:gradient}.
The value of $\xi_{<}(p)$ is determined by the underlying space topology and can thus be measured from a standard lattice using an appropriate estimation for $p$ in the $0<\rho<\rho_c$ region.
From Eq. (\ref{eq:p}), $p$ is not constant and an exact solution for $\xi_<$ would require treating the full gradient percolation problem~\cite{sapoval}.
In this work, for simplicity we take $\bar{p}$ which is the average of $p$ over the region of interest.
Additionally, the removal of the hole causes secondary damage due to dependencies in the annulus and the concentration of the gradient is decreased by a factor of $g$ which we measure numerically and which varies monotonically from $0.85$ to $0.89$ as a function of $r$.
We can thereby estimate $\bar{p} \approx  g(r) \int_0^{\rho_c} p(\rho) d\rho $.
We evaluate $\xi_{<}$ following \cite{bunde1991fractals} as:
\begin{equation}
 \xi_{<}^2 = \frac{1}{N_p} \sum _{(i,j)} \vert \mathbf{r}_i - \mathbf{r}_j \vert ^2
\end{equation}
where $(i,j)$ refers to nodes $i$ and $j$ which are in the same connected component and $N_p$ is the total number of such pairs of nodes.  This leads to a self-consistent condition for hole propagation:
\begin{equation}\label{eq:selfcons}
 \xi_{<} < \rho_c
\end{equation}
both sides are functions of $r , r_h$ and $p_s$.  Using these considerations, we can predict $r_h^c$ for every set of $(k,r)$ parameters with close agreement to the numerical results, see Fig.~\ref{simANDmodel}.

% The result of the depenedency links is that as their maximal length $r$ increases, a larger hole is needed to initiate a spreading cascade.

Everything about the scenario described above is local: nodes in $A$ and $B$ can have dependency links only up
to length $r$, the connectivity links in $A$ and $B$ are tied to an underlying lattice structure with characteristic length of one and the attack is restricted to a hole of radius $r_h$.
However, for a wide range of system parameters, this leads to a catastrophic cascade which destroys the entire system.
In fact, the localization of dependency opens the door for the spreading phenomenon that characterizes such a collapse.
When a hole of radius $r_h$ is removed from $A$, the nodes in $B$ that depended on them must be within a distance $r$ of the hole.
Thus the secondary damage is highly concentrated around the edge of the hole, leading to the creation of a damage front which propagates outwards from step to step.
If $r\rightarrow\infty$ or $r\rightarrow 0$, this weakness would not exist because the secondary damage would be spread out uniformly or remain in place, respectively.

In Summary, we find that paradoxically, the highly localized topology of embedded interdependent networks enables relatively small attacks to cause global damage. Given the low average degree of the power grid~\cite{hines-chaos2010,hines-proceedings2010} and its evidence of self-organizing criticality~\cite{carreras-ieee2004,dobson-chaos2007}, we anticipate high susceptibility even to relatively small local attacks.
%\section{\label{sec:Results}Results}

We acknowledge the LINC and MULTIPLEX (EU-FET project 317532) projects, the Deutsche Forschungsgemeinschaft (DFG), the Israel Science Foundation, ONR and DTRA for financial support.

%
% \begin{figure}%[ht*]
%   \begin{center}
%   \includegraphics[width = 8cm,height = 5cm]{r_min_vs_p_sim_rmax_200_iso_xi_c_0.15.eps}
%
%   \end{center}
%   % Requires \usepackage{graphicx}
%   %\includegraphics[]{}\\
%   \caption{\textbf{
%    a}}
%   \label{as}
% \end{figure}
%
% \begin{figure}[ht*]
%   \begin{center}
%   \includegraphics[width = 8cm,height = 5cm]{critical_hole_vs_r_min_sim_rmax_200_iso_xi_c_0.15.eps}
%
%   \end{center}
%   % Requires \usepackage{graphicx}
%   %\includegraphics[]{}\\
%   \caption{\textbf{s
%    }}
%   \label{asqw}
% \end{figure}

%\begin{figure}[ht*]
%  \begin{center}
%  \includegraphics[width = 8cm,height = 5cm]{contour_lines_p_vs_system_r.eps}

%  \end{center}
%  % Requires \usepackage{graphicx}
%  %\includegraphics[]{}\\
%  \caption{\textbf{s
%   }}
%  \label{as}
%\end{figure}

%\FloatBarrier

\bibliography{NoN}
% \bibliography{correlation_networks}
\end{document}